\documentclass[12pt]{article}
\usepackage[utf8]{inputenc}
\usepackage{booktabs}

\usepackage{newtxtext,newtxmath}

\usepackage{graphicx}

\usepackage[letterpaper,margin=1in]{geometry}

\renewenvironment{abstract}
{\quotation}
{\endquotation}

\date{}

\makeatletter
\renewcommand{\fnum@figure}{\textbf{Figure \thefigure}}
\renewcommand{\fnum@table}{\textbf{Table \thetable}}
\makeatother

\usepackage{scicite}

\usepackage{url}

\def\scititle{
	2D MoS$_2$/Au interfaces for enhanced opto-electronic response with sub-bandgap photons
}
\title{\bfseries \boldmath \scititle}

\author{
	J. Wu$^{1,2,3}$, W. Huang$^{1,3,4}$, A. Ali$^{1,3}$, Y. Chen$^{3,5,6}$, E. Stavrou$^{3,5,6}$, Z. Xie$^{3,5,7}$,\and J. Zhong$^{2,8,9}$, K. J. Karki$^{1,3\ast}$\and
	\small $^1$ Department of Physics, Guangdong Technion-Israel Institute of Technology, Shantou 515063, China	\and
	\small $^2$ State Key Laboratory of Quantum Functional Materials, Department of Materials Science and Engineering, \and \small Southern University of Science and Technology, Shenzhen 518055, China \and
	\small $^3$ Guangdong Provincial Key Laboratory of Materials and Technologies for Energy Conversion, \and \small Guangdong Technion-Israel Institute of Technology, Shantou 515063, China \and
	\small $^4$  Schulich Faculty of Chemistry and Solid State Institute, Technion-Israel Institute of Technology, \and \small Haifa 3200003, Israel	\and
	\small $^5$ Department of Material Science and Engineering, Guangdong Technion-Israel Institute of Technology, \and \small Shantou 515063, China	\and
	\small $^6$ Department of Materials Science and Engineering, Technion-Israel Institute of Technology, \and \small Haifa 3200003, Israel\and
	\small $^7$ Quantum Science Center of Guangdong-Hong Kong-Macao Greater Bay Area (Guangdong), \and \small Shenzhen-Hong Kong International Science and Technology Park, No. 3 Binglang Road, \and \small Futian District, Shenzhen, Guangdong 518000, China	\and
	\small $^8$ Guangdong Provincial Key Laboratory of Sustainable Biomimetic Materials and Green Energy, \and \small Southern University of Science and Technology, Shenzhen 518055, China\and
	\small $^9$ Institute of Innovative Materials, Southern University of Science and Technology, Shenzhen 518055, China \and
	\small$^\ast$Corresponding author. Email: khadga.karki@gtiit.edu.cn
}

\begin{document} 
	
	\maketitle



\begin{abstract}
Monolayer MoS$_2$ is a direct band gap semiconductor with potential applications in optoelectronics and photonics. MoS$_2$ also has a large optical nonlinearity. However, the atomic thickness of the monolayer limits the strength of the measured functional signals, such as the photocurrent or photoluminescence,  in optoelectronic devices. Here, we show that photocurrent in monolayer MoS$_2$ can be induced by sub-band gap photons by depositing Au nanoparticles on it. In this system, the nonlinear light-matter interaction in Au nanoparticles enhanced by the localized surface plasmons results in the generation of supercontinuum, which is reabsorbed by MoS$_2$ due to efficient resonant energy transfer. Au nanoparticle assisted photocurrent is more than an order of magnitude larger than two-photon photocurrent in monolayer MoS$_2$. Optimization of the shape, size and composition of the nanoparticle has the potential to enhance the photocurrent significantly with the prospect of applications in the detection of NIR photons, and related technologies including optical telecommunication.

\end{abstract}


\section{Introduction}

The discovery of two-dimensional (2D) materials has ushered in a new era in condensed matter physics and materials science, offering unprecedented opportunities for exploring quantum confinement effects and designing ultra-thin optoelectronic devices.\cite{Novoselov2004,Xia2014} Among these materials, transition metal dichalcogenides (TMDCs), and in particular molybdenum disulfide (MoS\textsubscript{2}), has emerged as a promising candidate due to their layer-dependent electronic properties, strong light-matter interactions, and exceptional mechanical flexibility.\cite{Wang2012,Radisavljevic2011} Unlike its bulk counterpart, monolayer MoS\textsubscript{2} exhibits a direct band gap of 1.90 eV, enabling efficient photoluminescence (PL) and photocurrent (PC), thereby making it ideal for applications in photodetectors, light-emitting diodes, and valleytronic devices.\cite{Yu2016,Choi2012}


The non-centrosymmetric crystal structure of monolayer MoS\textsubscript{2} governs its nonlinear optical properties, resulting in a large second-order nonlinear susceptibility ($\chi^{(2)} \approx 10^2$ pm/V),\cite{Heinz2017} which enables efficient second harmonic generation (SHG), sum-frequency generation (SFG), and optical rectification.  This strong nonlinear response makes MoS\textsubscript{2} an attractive material for nonlinear optical applications.\cite{Wu2018,Jadidi2016} However, the low absorption efficiency of a monolayer under typical excitation intensities results in a weak two-photon photoluminescence (TPL) and two-photon photocurrent (TPC) signals, limiting utility in nonlinear response based optoelectronics that include ultrafast optoelectric switches, subbandgap photodetectors, etc.\cite{Cheng2011} Interfacing the monolayer of MoS\textsubscript{2} with plasmonic systems provides a compelling solution to this problem, as the nonlinear response from the hybrid system can significantly increase the excitation, boosting external response without causing material damage.\cite{Butun2015,Guo2015}

In this work, we investigate the 
nonlinear optical phenomena in Au-nanoparticles deposited on monolayer MoS\textsubscript{2}, focusing on TPL and TPC. The integration of plasmonic nanoparticles with 2D materials can be versatile strategy for overcoming the intrinsic limitations of atomically thin systems. By harnessing the effects of nonlinear response from the plasmonic system, and long-range energy and charge transfer at the interface, one can unlock the full potential of nonlinear optical phenomena in the hybrid system, enabling novel applications in high-sensitivity photodetection, ultrafast optical switching, and nonlinear on-chip light sources.\cite{Oulton2009,Kauranen2012}
Our results reveal that the interplay between localized surface plasmons (LSPs) enhanced nonlinear response from Au-nanoparticles, in particular supercontinuum generation,  and excitonic transitions in MoS\textsubscript{2} induced by the supercontinuum leads to a 15-fold boost in PC compared to the intrinsic TPC in monolayer MoS$_2$ under femtosecond laser excitation. 
 This work advances our fundamental understanding of Au-enhanced nonlinear optics and provides a road map for designing next-generation optoelectronic devices that leverage the unique properties of 2D materials.\cite{Lin2023} As the field continues to evolve, the synergy between plasmonic structures and 2D materials is poised to drive breakthroughs in photonics, quantum technologies, and beyond.

\section{Results and discussion}
\begin{figure}
  \includegraphics[width=0.9\linewidth]{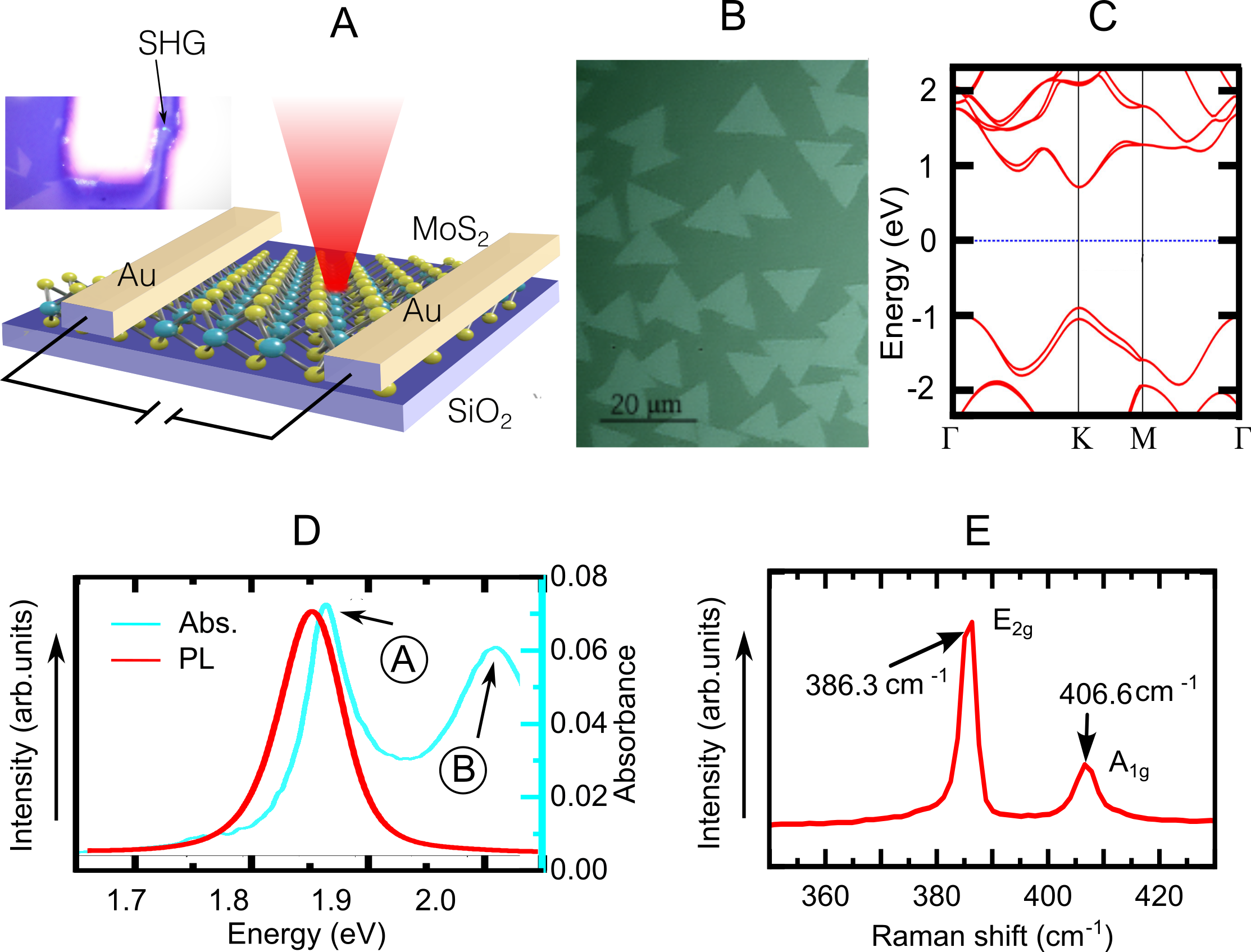}
  \caption{(A) Schematic of a device used in the experiments showing a monolayer MoS\textsubscript{2} in contact with Au electrodes. The inset shows a photo of a device excited by a pulsed laser at 1030 nm. SHG from a flake of MoS$_2$ is indicated. (B) Image of flakes of MoS$_2$ on a sapphire substrate onto which the electrodes are deposited. (C) Theoretically calculated band structure of single-layer MoS$_2$ (adapted from Ref.\cite{GERBI2025}). (D) Absorption (blue) and PL (red) spectra of the sample, showing a red shift in the PL with respect to the absorption. (E) Raman spectrum of the sample confirming that sample is composed of monolayer MoS$_2$.}
  \label{fig1}
\end{figure}

A schematic of the device used for optical and electrical measurements is shown in Figure~\ref{fig1}(A). It consists of flakes of MoS$_2$ on a sapphire substrate. Au electrodes are deposited on the flakes allowing optical access in between the electrodes. An image of the device emitting SHG when irradiated with femtosecond pulses is shown in the inset and a micrograph of the sample is presented in Figure~\ref{fig1}(B). The triangular flakes of MoS$_2$, which are seen as lighter regions in the micrograph, are smaller than 50 $\mu$m in length. The electronic structure of MoS$_2$ is sensitive to the number of layers. Figure~\ref{fig1}(C) shows a representative band structure of monolayer MoS$_2$ computed using density functional theory.~\cite{GERBI2025} The computed bandgaps show significant variation from 1.6 to 2.4 eV depending on the methods used.~\cite{GERBI2025,LAMBRECHT2012} Nevertheless, most of them elicit a splitting of the valence band due to spin-orbit coupling, resulting in two excitonic transitions, referred to as A and B excitons, which are observed in the absorption spectra shown in Figure~\ref{fig1}(D). The bandgap obtained experimentally is 1.9 eV.~\cite{HEINZ2010} As the splitting of the valence band is also present in multilayer MoS$_2$, the number of layers in the flakes cannot be inferred from the absorption spectrum. Similarly, one cannot quantify the layer composition from the PL spectrum as it is also observed in flakes with few layers. More importantly, the PL from MoS$_2$ depends on the excitation wavelength.~\cite{GIES2015,KAXIRAS2016} The spectrum obtained by excitation at 450 nm, shown in Figure~\ref{fig1}(D), has a prominent peak at 1.85 eV. The PL is dominated by the recombination of excitons at the band edge. Among different spectroscopic methods, Raman spectroscopy has proven to be sensitive to the Van-der Waals interactions between the layers. Hence, we have used it to identify the number of layers in the flakes. The positions of two characteristic Raman modes, namely the $A_{1g}$ mode associated with the out-of-plane vibration of sulfur atoms and the $E_{2g}$ mode related to the in-plane vibration of Mo and sulfur atoms, are sensitive to the number of layers.~\cite{Li2012} The Raman spectrum of the specimen in Figure~\ref{fig1}(E) clearly exhibits two Raman active modes: the $E_{2g}$ mode at 386.3 cm$^{-1}$ and the $A_{1g}$ mode at 406.6 cm$^{-1}$. The frequency separation between the two peaks is 20.3 cm$^{-1}$, which is a typical value for a monolayer of MoS$_2$.~\cite{Li2012,yu2013controlled}  The layer composition has also been confirmed by the measurement of height of the flakes using an AFM (see Experimental section, Figure~\ref{fig6}(E)).

\begin{figure}[htb]
  \includegraphics[width=0.9\linewidth]{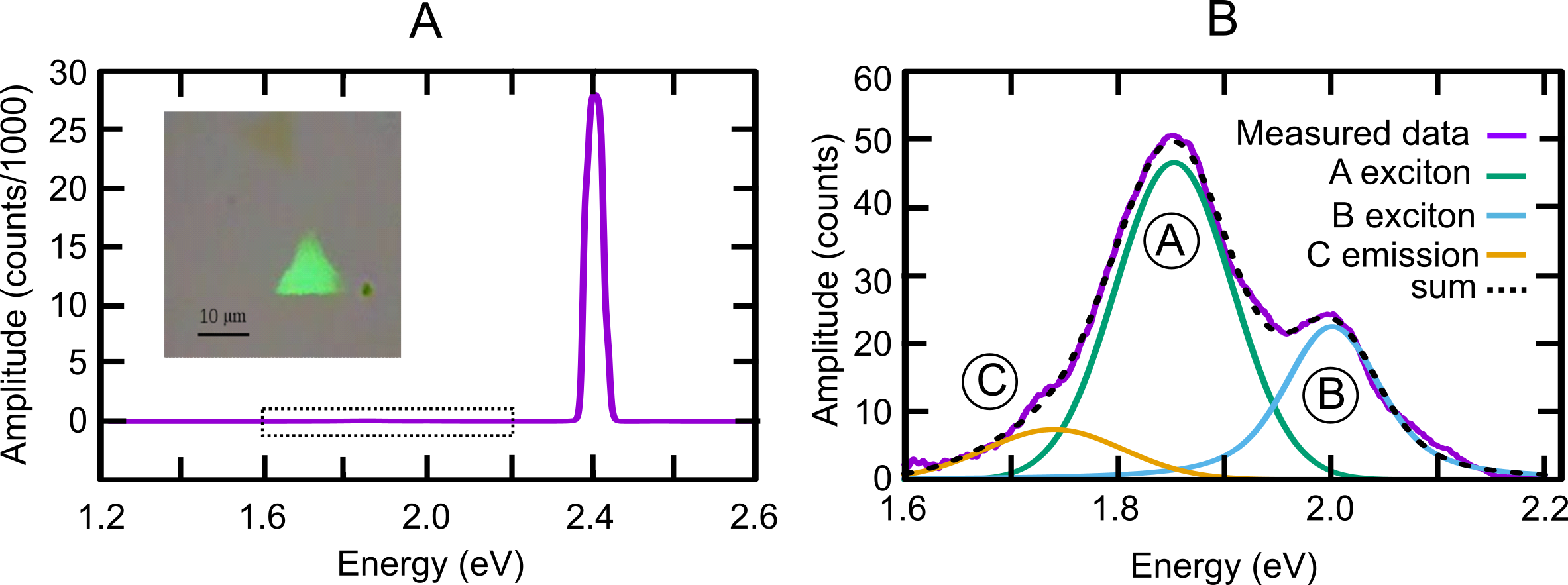}
  \caption{(A) Spectrum of the emission from monolayer MoS$_2$ induced by femtosecond pulses at 1030 nm (photon energy of 1.2 eV). A micrograph of the green emission visible to the naked eye is shown in the inset. The emission is dominated by SHG. A weaker emission, present in the dotted region, is shown in detail in (B). Three peaks are evident in the emission spectrum. }
  \label{fig2}
\end{figure}

The excitation of MoS$_2$ with femtosecond pulses at 1030 nm with an equivalent photon energy of 1.2 eV gives rise to different types of nonlinear responses. The spectrum of  optical nonlinear response, shown in Figure~\ref{fig2}(A), has a distinct contribution from the SHG at 2.4 eV, which is also visible as green emission as shown in the inset. A closer analysis reveals additional emission peaks due to TPL indicated by dotted box in the region between 1.6 and 2.2 eV.  The comparatively lower detection of PL can be attributed to two causes. First, because it is emitted uniformly in all directions, only a fraction of the PL is collected by the microscope objective. The estimated collection efficiency of an objective having a working distance of 16 mm and pupil diameter of 14 mm (Nikon CFI Plan Fluor, 10X objective) is approximately 4\%. However, the collection efficiency of the SHG is 50\%, regardless of whether it is collected in the forward or epi-direction. After correcting for the collection efficiency, we find that the SHG yield is still an order of magnitude larger than TPL. The details of the comparison are provided in the Supplementary Information.

The TPL spectrum shown in Figure~\ref{fig2}(B) has three peaks, which differs from the spectrum of  PL excited by a CW laser at 450 nm.  It is well known that the excitation wavelength affects the PL spectrum's features.~\cite{GIES2015,KAXIRAS2016} Specifically, excitation at 450 nm raises electrons to energy levels well above the band gap, causing the ensuing relaxation to produce a thermally equilibrated population at the band edge, which is equivalent to a dominant population of A excitons.
However, the excitation by two-photon absorption (TPA) is energetically comparable to the excitation by photons at 515 nm, which, being quasi-resonant with both excitonic levels, populates both of them. As shown in the figure, this results in two peaks in the PL spectrum that represent the recombination of A and B excitons. ~\cite{GIES2015} 
The third peak, indicated as C, has also been observed in such excitation. However, its origin is debated as either arising from the recombination of trapped excitons~\cite{KAXIRAS2016} or from the recombination of trions.
~\cite{SWAN2017,QU2021,DHARA2022} The parameters of the peaks are tabulated in Table~\ref{table1}. Based on the area under the peaks, A excitons contribute the most (60\%), followed by B excitons (29\%) and C emission (12\%).

\begin{table} [htbp]
 \caption{Fitting parameters of the peaks in the TPL spectrum shown in Figure~\ref{fig2}(B)}
 \begin{center}
  \begin{tabular}[htbp]{@{}lllll@{}}
    \hline
    Peak & Center (eV) & Amplitude & FWHM (eV) & Fraction of the total PL  \\
    \hline
    A  & 1.85  & 46.5 & 0.13 eV & 0.59  \\
    B  & 2.0  & 22.5 & 0.11 & 0.29  \\
    C  & 1.74 & 7.4 & 0.15 & 0.12 \\
    \hline
  \end{tabular}
  \end{center}
  \label{table1}
\end{table}


Next, we analyze the intensity dependence of PL, PC, SHG, TPL and TPC. To accurately quantify the dependence, a method based on modulated excitation and detection of the additional modulations in the different responses is used.~\cite{karki2015phase} In this method, the intensity of a laser beam is modulated at a single frequency, $\phi$, which is set to 2 kHz in the measurements. The modulation frequency can be chosen in the range of few kHz (see the materials and methods section for details) based on the response time of the measured system and spectrum of the noise. One typically chooses the modulation frequency to be less than the bandwidth of the complete measurement system including photodectors, preamplifiers and the data acquisition system. The intensity of the modulated beam can be expressed as $I(t) \propto A+B \cos(\phi t)$, where the ratio of $A$ and $B$ give the depth of the modulation. Linear response, such as PL and PC due to single photon excitation, modulate at the same frequency $\phi$ but nonlinear responses show additional modulations at multiple harmonics. 

\begin{figure}[htb]
	\includegraphics[width=0.9\linewidth]{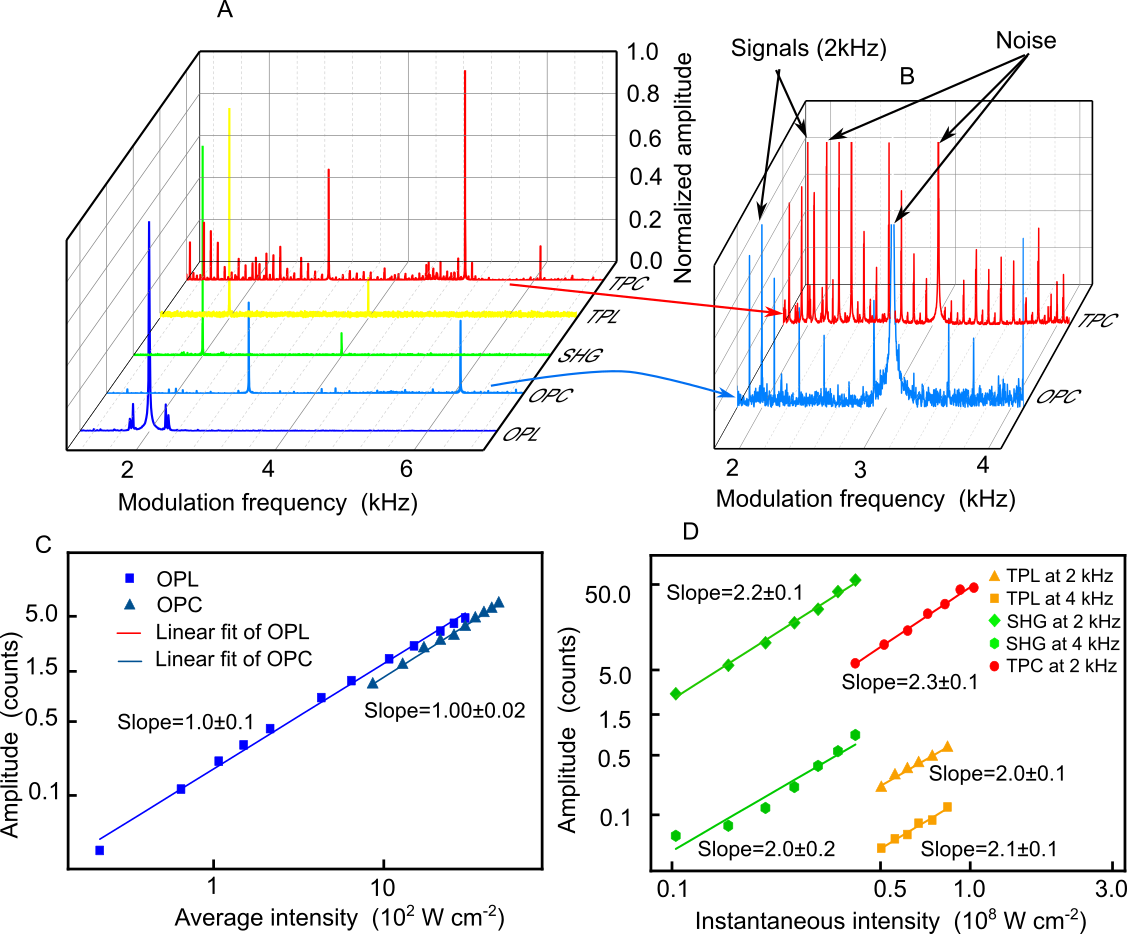}
	\caption{(A) Fast Fourier Transforms (FFT) of different measured signals induced by excitation modulated at 2 kHz. (B) Comparison of signal and noise in the FFT of PC. (C) Intensity dependence of single photon PC and PL, and (D) intensity dependence of different nonlinear signals.  }
	\label{fig3}
\end{figure}

Figure~\ref{fig3}(A) shows the Fast Fourier transforms (FFT) of the different responses. The PL and PC obtained from the monolayer MoS$_2$ using a CW laser at 450 nm show modulations at $\phi=2$ kHz, which matches with the modulation frequency of the excitation beam. As can be seen in (B), the PC is dominated by large electronic interference at multiple frequencies. Nevertheless, the signal at 2 kHz can be isolated as the electronic noise at this frequency is minimal. The lack of modulation at harmonics of the 2 kHz indicates that the one-photon PL (OPL) and PC (OPC) induced by the excitation at 450 nm are linear. In contrast, SHG and TPL show distinct modulations at 2 and 4 kHz indicating that the signals are nonlinear. More importantly, modulations at higher order harmonics, for example, at 6 and 8 kHz, are absent. This has important implications for the dynamics of free electrons, holes, excitons, and other quasiparticles that contribute to the emissions. First, as SHG is an instantaneous coherent response from the sample, it is not affected by the dynamics of excited particles. Being a second order response with a quadratic dependence on the excitation, its intensity is given by $I_{SHG}(t) \propto I^2(t) \propto 4 A B  \cos(\phi t)+B^2 \cos(2\phi t)$,~\cite{karki2015phase} showing modulations exactly at 2 and 4 kHz, which agree with the measurements. On the other hand, as TPL and TPC are not instantaneous, the ensuing dynamics of the excited particles alter the frequency response. In this case, the initial populations of the excited electrons, $n_e$, and holes, $n_h$, are given by TPA: 
\begin{equation}
	n_e(t)=n_h(t)=T_A(t)  \propto I^2(t) \propto 4 A B  \cos(\phi t)+B^2 \cos(2\phi t),
\end{equation}
which also modulate at $\phi$ and $2\phi$. However, the measured PL and PC need not depend linearly on $n_e$ or $n_h$. If the PL is due to the recombination of free carriers or the excitons formed by the nongeminate combination of electrons and holes then  $I_{PL}(t)\propto n_e(t) n_h(t) \propto T_A^2(t)$. In these cases, modulations at $\phi,\, 2\phi,\, 3\phi$ and $4\phi$ are observed.~\cite{KARKI2019} The observation of modulations only at $\phi=2$ kHz and $2\phi=4$ kHz indicates that only the excitons formed by geminate combinations of electrons and holes contribute to the PL, implying that the excitons are formed at extremely short time scales such that an excited pair of electron and hole do not have time to combine nongeminately with electrons and holes generated at other locations. It is important to discuss the implications of modulation frequencies in TPL on the origin of C emission shown in Figure~\ref{fig2}(B). If the emission is from the trapped excitons that are produced by geminate combination of electrons and holes, then TPL from them modulates at the two frequencies. However, if the emission is from trions, the ensuing dynamics is nonlinear, resulting in modulations at higher order harmonics, which is not consistent with our observation. Nevertheless, it has been argued that in some special cases, the decay of trion population can be linear. This points to a need for further study on their dynamics.

Figure~\ref{fig3}(B) shows the FFT of PC in the frequency range of 1.8 to 4.3 kHz where the signal at 2 kHz is evident but signals at its harmonics are not discernible. Although this is expected for the excitation at 450 nm, the lack of clear signal at 4 kHz in TPC is due to the large background noise. One simple approach to minimize the interference from the noise would be to modulate the excitation beam at higher frequency, where the noise is less. As can be seen in Figure~\ref{fig3}(A), the noise above 6 kHz is substantially reduced. However, in the measurement of PC, the bandwidth of the amplifier and amplification factor need to be considered, which limits the highest modulation frequency that can be used in our measurements.

 The dependence of OPL and OPC on the excitation intensity is shown in Figure~\ref{fig3}(C). Both the signals increase linearly in agreement with the response in the frequency domain shown in (A). On the other hand, SHG, TPL and TPC scale quadratically with the excitation intensity (see Figure~\ref{fig3}(D)) providing further evidence of the second order nonlinear response as the origin of SHG, and TPA followed by linear relaxation as the origin of TPL and TPC.
 
\begin{figure}[htb]
	\centering
	\includegraphics[width=0.9\linewidth]{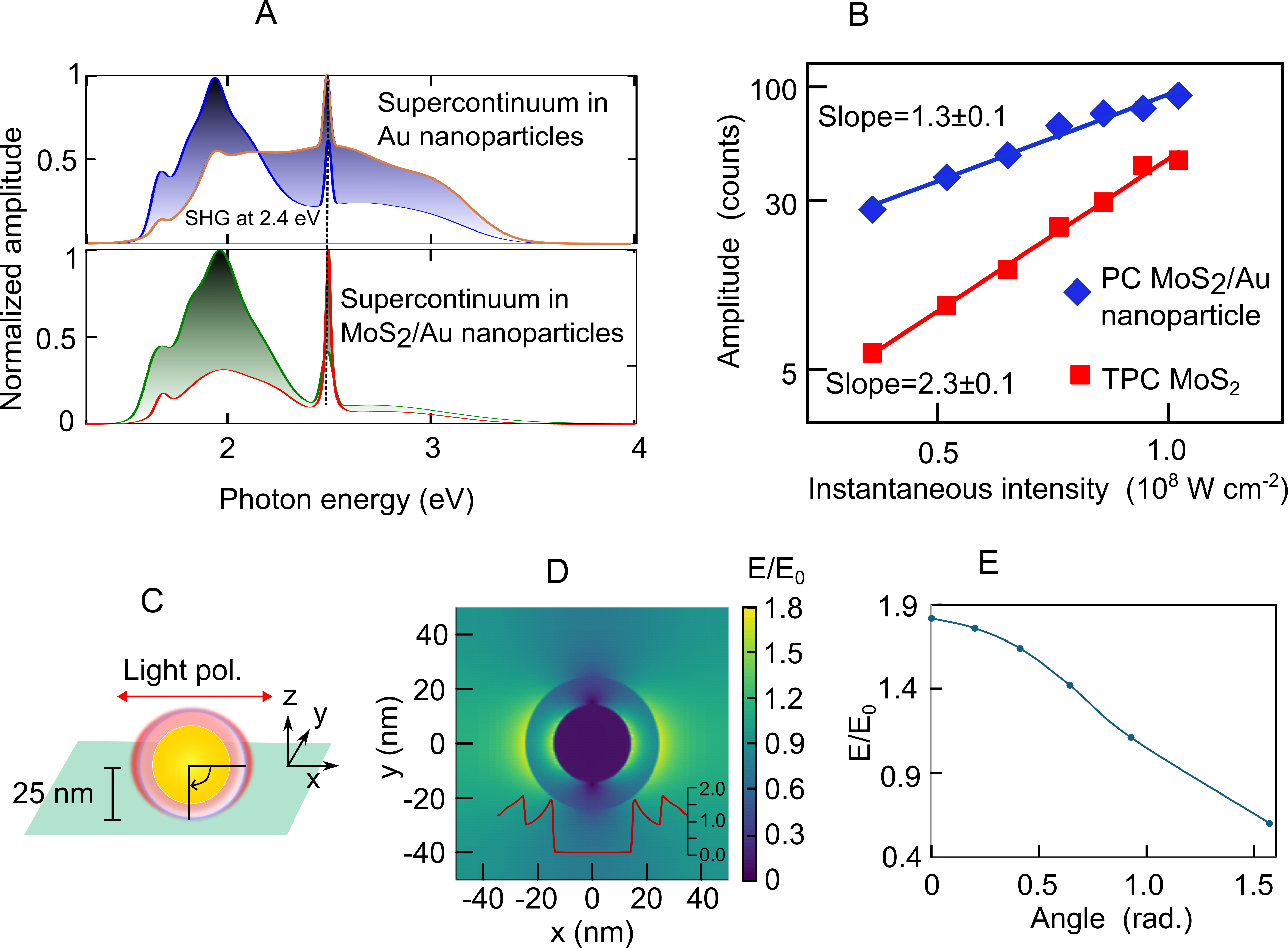}
	\caption{(A) Top: Supercontinuum from Au nanoparticles on sapphire substrate irradiated by femtosecond pulses at 1030 nm. Bottom: Supercontinuum from Au nanoparticles on MoS$_2$. The spectra vary at different locations on the samples. (B) Intensity dependence of PC from MoS$_2$ and MoS$_2$ with Au nanoparticles. (C) Illustration of Au nanoparticles on MoS$_2$ showing the polarization of the light, regions around the particle where electric fields are computed using FDTD. (D) Electric field around the circumference of the nanoparticle on the plane parallel to MoS$_2$/substrate and passing through the center of the particle. (E) Electric field on the surface of the nanoparticle as a function of angle (0 to $\pi/2$) as the point is moved from the center of the particle to the base where it is in contact with MoS$_2$. }
	\label{fig4}
\end{figure}

 Next we investigate the optical response of 2D-MoS$_2$/Au nanoparticles under pulsed excitation. Au/SiO$_2$ core-shell nanoparticles are used in the measurements to avoid a direct contact between Au and MoS$_2$. The Au core of the nanoparticles have an average diameter of about 26 nm while the thickness of the shell is about 8 nm. The particles have the resonance peak of LSPs at 522 nm (the absorption spectrum is given in the Experimental section). Previous studies have shown that the LSPs in Au films elicit strong nonlinear response when excited by femtosecond pulses resulting in the emission of a super-continuum whose spectra extend from UV to NIR wavelengths.~\cite{MATHUR2013,KARKI2018,ZAYATS2016} One expects similar response from Au nanoparticles as they also support LSPs. A comparison of emission from Au nanoparticles on sapphire substrate (top) with MoS$_2$/Au nanoparticles hybrid system (bottom) is shown in Figure~\ref{fig4}(A). In general, the measured spectrum of supercontinuum from Au nanoparticles in sapphire extends up to 3.5 eV (~350 nm) in the UV and 1.6 eV (~775 nm) in the NIR. It is important to note that the spectrum in the NIR is clipped by short-pass filter at 750 nm that is used to filter out the excitation beam. The actual spectrum of the supercontinuum extends to wavelengths longer than 1030 nm. The obvious sharp peak at 2.4 eV (515 nm) is due to the SHG from the nanoparticles. More importantly, a pronounced variation in the spectra with the position in the sample is observed due to the variation in the distribution of the nanoparticles. The emission from the hybrid system of Au nanoparticles on monolayer MoS$_2$ is similar. Because of the large variation in the spectra in both the systems, quantification of enhancement or suppression of the PL from MoS$_2$ by the Au nanoparticles does not provide statistically meaningful result. To overcome this difficulty, we have compared the PC from the two systems. 
   
Figure~\ref{fig4}(B) shows the PC at 2 kHz from MoS$_2$ and MoS$_2$ with Au nanoparticles at different excitation intensities. PC from the hybrid system is consistently higher at all the measured intensities. However, the intensity dependence of the PC from the hybrid system is subquadratic, indicating that the dominant contribution is not from TPA in MoS$_2$. To investigate the mechanism of the enhancement, we have considered three possibilities: (i) enhancement of the local electric field by LSPs in the nanoparticles, (ii) tunneling of the electrons excited in the Au to MoS$_2$, and (iii) long-range energy transfer from the nanoparticles to MoS$_2$.

First we discuss the enhancement of the electric field by the nanoparticles. Figure~\ref{fig4}(C) shows the model of a core-shell Au/SiO$_2$ nanoparticle used to simulate the electric field around the nanoparticle by using  Finite-Difference Time-Domain (FDTD) methods when the nanoparticle is irradiated by pulsed laser. The parameters of the laser pulses match the experiments (see the Supplementary Information for the details of the simulations). The polarization of the beam is parallel to the plane of the substrate on which MoS$_2$ is placed (x-y plane). The field distribution on the x-y plane that passes through the center of the particle is shown in (D). The field is enhanced parallel to the polarization of the light on the Au/SiO$_2$ and SiO$_2$/air interfaces at y=0, while it is suppressed in the orthogonal direction. The maximum enhancement factor at Au/SiO$_2$ is  1.8, which is rather modest owing to the nonresonant excitation of LSPs in Au nanoparticle.  Similar enhancement factor attributable to dielectric enhancement is present at the SiO$_2$/air interface.~\cite{ALBELLA2019} However, the enhanced fields are not in the vicinity of the surface of MoS$_2$ to directly increase the absorption. Figure~\ref{fig4}(E) shows the ratios of the amplitude of the near-field to the average amplitude at the far-field along the circumference of the nanoparticle as the function of the angle indicated in (C). The ratio decreases from 1.8 to 0.6. In fact, at the base of the nanoparticle, where it is in contact with MoS$_2$ layer, the field is suppressed. 

\begin{figure}[htb]
	\centering
	\includegraphics[width=0.9\linewidth]{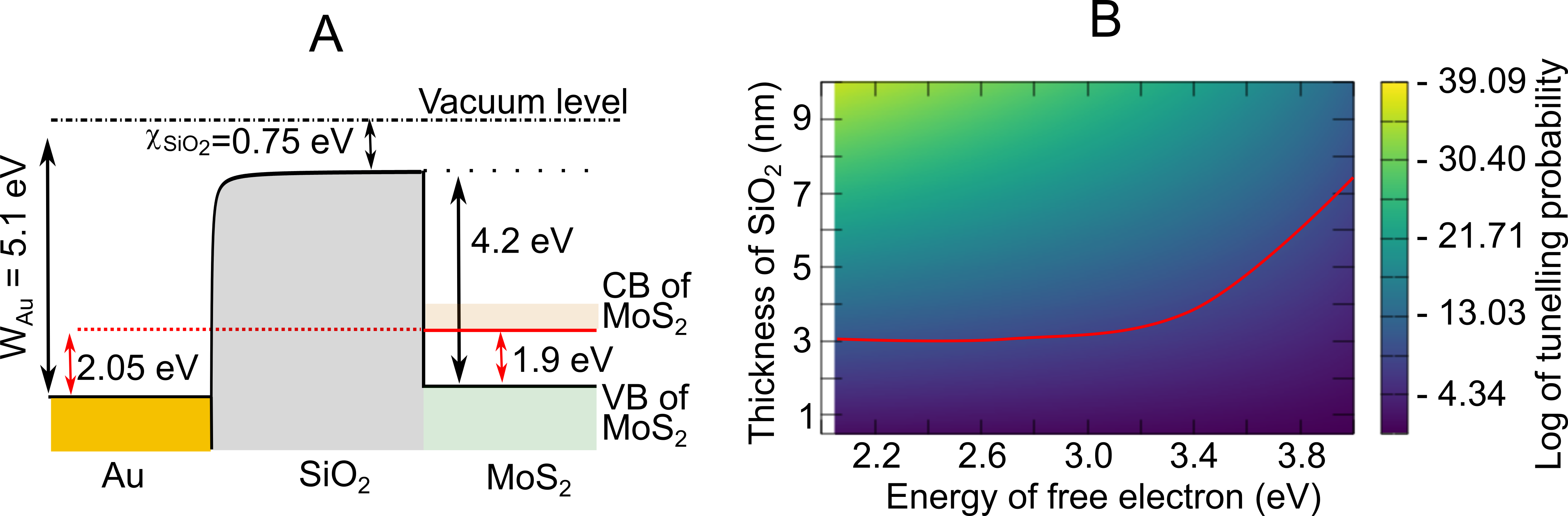}
	\caption{(A) Band alignment of Au/SiO$_2$/MoS$_2$ heterostructure and (B) tunneling probability of electrons from Au to MoS$_2$ as a function of the energy of hot electrons and thickness of SiO$_2$.  }
	\label{fig5}
\end{figure}

Next, we discuss the possibility of excited free electrons tunneling from Au to MoS$_2$. Figure~\ref{fig5}(A) shows the band alignment of Au/SiO$_2$/MoS$_2$ system based on the work function of Au ($W_\textrm{Au}=5.1$ eV), electron affinity of SiO$_2$ ($\chi_{\textrm{SiO}_2}$=0.7 5 eV),~\cite{MIYAZAKI2016}, and the valence band of MoS$_2$ relative to the conduction band of SiO$_2$ (4.2 eV).~\cite{STEMANS2015} According to the band diagram, only the excited electrons with energies $>$ 2.05 eV that can tunnel to the conduction band of MoS$_2$ contribute to photocurrent. The potential at Au/SiO$_2$ interface is given by 

\begin{equation}\label{potential}
	V(x)=(W_{\textrm{Au}}-\chi_{\textrm{SiO}_2})-\frac{e^2}{16\pi\epsilon_0 \epsilon_{\textrm{SiO}_2} x},
\end{equation}    
where $e$ is the charge of an electron, $\epsilon_0$ is vacuum permittivity, $\epsilon_{\textrm{SiO}_2}$ is the relative permittivity of SiO$_2$ and $x$ is the distance from the Au interface. The last term in Equ.(\ref{potential}) is due to the image charge. The tunneling probability for an electron with energy $E$ and effective mass ($m_e^*=0.3 m_o$) in SiO$_2$~\cite{HENKER2007} is given by 
\begin{equation}\label{tunelling}
	T_t = \exp\left[-2\frac{\sqrt[]{2 m_e^*}}{\hbar} \int_{0}^d \sqrt[]{V(x)-E} \, dx\right].
\end{equation} 
  
Figure~\ref{fig5}(B) shows the tunneling probability of hot electrons from Au to MoS$_2$ as a function of the energy of free electrons in Au and the thickness of SiO$_2$. The red line separates regions with probabilities higher (lower region) or lower (upper region) than $10^{-9}$. In the lower region, the tunneling competes with resonant energy transfer processes, while it's contribution is negligible in the upper region. In particular, the tunneling of hot electrons through 8 nm thick silica shell is negligible for even the highest energies (3.5 eV) of the free electrons obtained from the emission spectrum in Figure~\ref{fig4}(A).

Finally, we discuss the possibility of energy transfer from Au nanoparticles to MoS$_2$. The distance of 8 nm from the Au nanoparticles to MoS$_2$ is within the length scale of F\"orster Resonance Energy Transfer (FRET). The rate of energy transfer ($k_{ET}$) from a point particle to a 2D material is inversely proportional to the fourth power of distance ($k_{ET}\propto 1/d^4$)  because of which significant transfer rates have been observed even at the distances of few tens of nanometers.~\cite{SEBASTIAN2009,WRACHTRUP2013} 
A peculiar feature of energy transfer in this system, which is markedly different from atomic or molecular systems, is that the emission and absorption are broadband. All the emission from Au nanoparticles above the bandgap of MoS$_2$ with $E_{ph} \geq 1.9$ eV can undergo FRET relaxation for which the standard methods of evaluation of its efficiency may not be adequate. Nevertheless, an estimation based on them is useful for comparison. It is usual to compute the FRET efficiency relative to the absorbed energy. However, in Au nanoparticles, a large fraction of absorbed energy dissipates nonradiatively yielding a rather low quantum yield of emission in the range of $10^{-6}$~\cite{CARUSO2004}. We estimate the FRET efficiency relative to number of photons emitted. The estimated efficiency, averaged over emission spectrum ($1.9 eV\leq E_{ph}\leq 3.5 eV$) is about 35\% (see Supplementary information for the details and ref.~\cite{KIM2024} for the dielectric function of monolayer MoS$_2$ and ref.~\cite{Mathur2021-ml} for the expression for the efficiency).  Thus, the efficiency of photons absorbed by MoS$_2$ via hot electrons in Au is in the order of $10^{-7}-10^{-8}$, which is dominant compared to electron transfer via tunneling when the thickness of SiO$_2$ is $> 3$ nm. The intensity dependence of PC in Figure~\ref{fig4}(B) also points to the contribution FRET in PC at low intensities. In particular, PC in Au/MoS$_2$ increases with an exponent of $1.3$, which indicates that the main contribution is not from TPA in MoS$_2$. On the other hand, generation of hot electrons in Au nanoparticles, and emission from them has complex dependence on the intensity, ranging from linear to highly nonlinear at low to high excitation intensities. The intensities used in our measurements correspond to low intensity regime, resulting in the sub-quadratic scaling of PC.

Photons at NIR wavelengths are widely used in telecommunication, nonlinear microscopy, and sensing.  Au-nanoparticle enhanced PC in MoS$_2$ by sub-bandgap photons can extend their use at these wavelengths. Although an order of magnitude enhancement compared to the direct nonlinear excitation is not enough for applications, further research on the influence of the size and shape of nano-particles has the prospect for dramatic enhancements. Similarly, here we have focused on energy transfer. But by controlling the thickness of SiO$_2$, it may be feabile to control the ratio of energy and charge transfer. 


\section{Conclusion}
To conclude, we have shown that core-shell Au nanoparticles emit broadband white light when excited by femtosecond pulses at 1030 nm at excitation intensities $<10^8$ W cm$^{-2}$. We have shown that the PC from MoS$_2$ can be enhanced by the efficient energy transfer from the Au nanoparticles even when the energy of the incident photons is less than the band gap. Our results elucidate a new mechanism to enhance functional signals in MoS$_2$ based devices by employing the plasmonic enhanced nonlinear response from Au nanoparticles. Specifically, PC in MoS$_2$ due to sub-bandgap photons at NIR wavelengths is highly attractive for extending its applications in telecommunications, photonics and optoelectronics. MoS$_2$/Au nanoparticles hybrid system serves as a platform to investigate the fundamentals of energy transfer between 0D- and 2D-materials.

\section{Experimental Section}
\subsection{Sample preparation and characterization}
\emph{Synthesis of Au nanoparticles.}
Au nanoparticles were synthesized by the seed mediated successive reduction method outlined in ref.~\cite{MULVANEY1996,NG2020}. Briefly, an ice-cold 5ml of 3 mM sodium borohydride (NaBH$_4$) solution obtained from Sinopharm (CAS\# 16940-66-2) was added dropwise to the pre-mixed KAuCl$_4$ (0.75 mM, 5 mL) (obtained from Macklin, CAS\# 13682-61-6) and sodium citrate (3.75 mM, 5 mL) solution (obtained from Macklin, CAS\# 6132-04-3). After 45 min of continuous reaction under atmospheric conditions, 5 nm Au naoparticle seeds were obtained. Then, under the action of continuous reduction by KAuCl$_4$ and ascorbic acid solution (obtained from Macklin, CAS\# 50-81-7), the 5 nm Au  seeds were grown into 26 nm particles, the whole synthesis process was proceeded with magnetic stirring. 8 nm thick silica shell was synthesized in two steps. First a shell of 5 nm was synthesized as follows. The concentrations of the solution of 26 nm Au nanoparticles were first adjusted to keep the total surface area of the Au nanoparticles at about $5 \times 10^{14}$ nm$^2$/mL. Then, 100 $\mu$L of 1 mM freshly prepared (3-aminopropyl) trimethoxysilane (APS) aqueous solution obtained from Sigma aldrich (CAS\# 13822-56-5) was added to 15 mL of the Au nanoparticle solution under vigorous magnetic stirring at room temperature. After mixing for 30 min, 2.5 mL of 0.54wt \% sodium silicate aqueous solution (pH = ca. 12) obtained from Sigma aldrich was added to the mixture and stirred for 3 minutes. The entire mixture was transferred to a 90 $^{\circ}$C water bath to accelerate the rate of silica coating. After vigorous stirring in a 90 $^{\circ}$C water bath for 1 hour, it was cooled to room temperature to obtain 5 nm thick silica shell-coated Au nanoparticles (pH = ca. 8.5). To purify the product, the reacted Au nanoparticle solution was centrifuged at 3000 rpm for 50 minutes, the precipitate was redispersed in an equal volume of water, and then centrifuged under the same conditions, and the process was repeated five times. The whole synthesis process was proceeded with magnetic stirring. Next, 6 mL of a solution of Au nanoparticles with a 5 nm thick silica shell was taken as seeds. First, the solvent of the 6 mL Au nanoparticle solution was replaced from water to an equal volume of 80\% ethanol obtained from Xilong (CAS\# 64-17-5). Under vigorous stirring, 900 $\mu$L of 28\% (v/v) ammonia solution obtained from Xilong (CAS\# 1336-21-6) was added to the solution to adjust the pH to about 9.0, and then 40 $\mu$L of 5\% Tetraethyl orthosilicate(TEOS) (v/v, in ethanol) obtained from Sigma aldrich (CAS\# 78-10-4) was added and stirred at room temperature for 3 minutes. The mixture was then transferred to a 90 $^{\circ}$C water bath, magnetically stirred for 30 minutes, and then cooled to room temperature. To purify the synthesized silica-shelled Au nanoparticles, the same steps above to isolate the Au nanoparticles was followed.

\emph{text}

\begin{figure}[htb]
	\centering
	\includegraphics[width=0.9\linewidth]{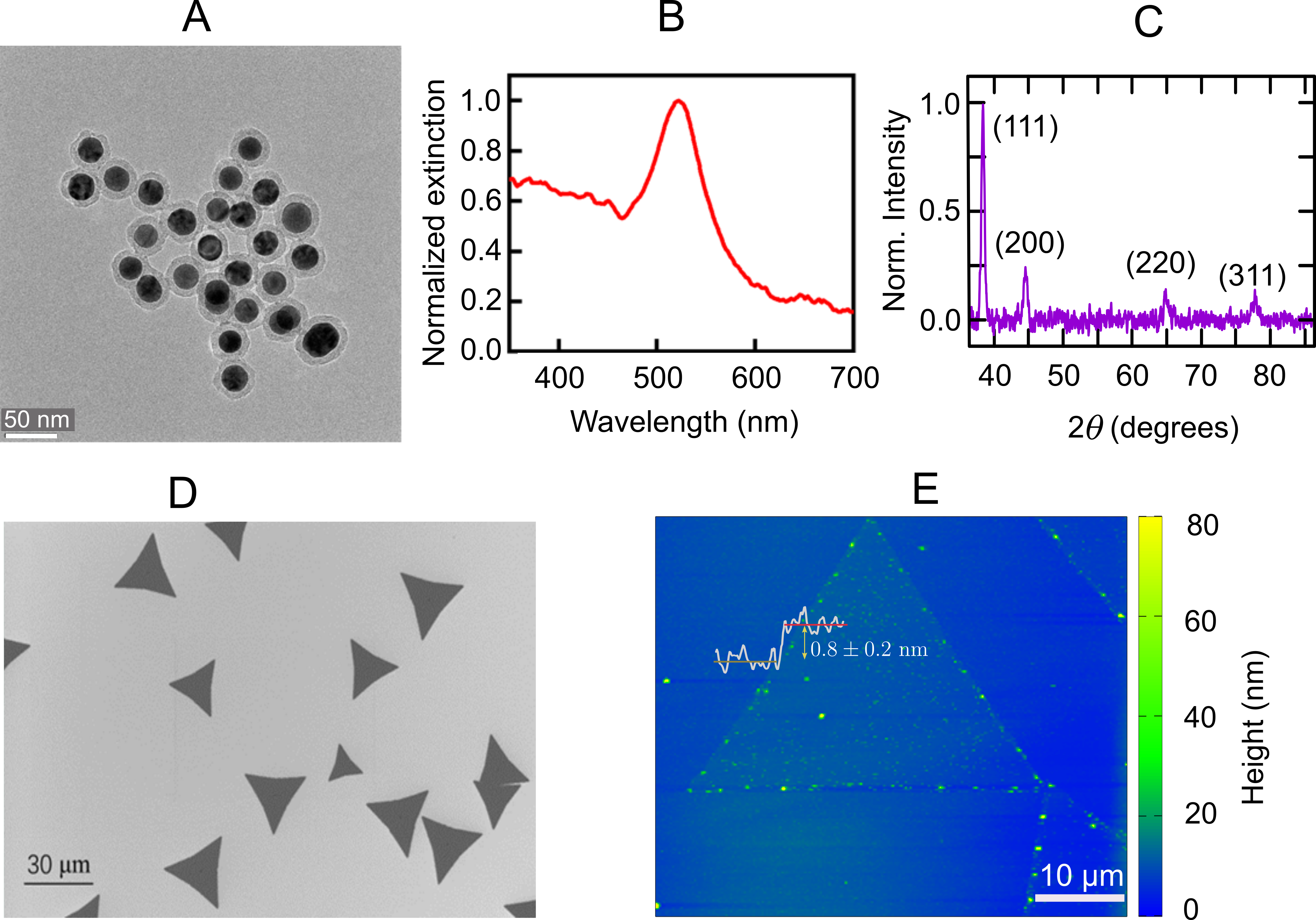}
	\caption{(A) TEM image of Au nanoparticles with silica shell. (B) UV-VIS absorption spectrum, and (C) powder X-ray diffraction  of the nanoparticles. (D) SEM, and (E) AFM images of MoS$_2$ flakes on silica.  }
	\label{fig6}
\end{figure}

\emph{Characterization of Au nanoparticles.} A transmission electron microscopy (TEM), Talos F200X G2 from ThermoFisher Scientific, was used to characterize the shape and size of the Au nanoparticles. Figure~\ref{fig6}(A) shows the TEM image of a cluster of nanoparticles. The Au cores and SiO$_2$ shells have distinctly different contrasts. The average diameter of the core is 26$\pm$1.6 nm, and the thickness of the shell is $8\pm0.7$ nm. A UV-vis absorption spectrometer was used to measure the extinction spectra of the nanoparticles. The extinction spectrum shown in Figure~\ref{fig6}(B) has a peak at 522 nm corresponding to the resonance of LSP. Powder XRD was used to test the crystallinity of Au. The XRD pattern shown in Figure~\ref{fig6}(C) has a major peak at 38.35$^{\circ}$, and other smaller peaks at 44.51$^{\circ}$, 64.76$^{\circ}$, and 77.73$^{\circ}$, which correspond to (111), (200), (220), and (311) Bragg reflection in good agreement with face centered cubic structure of metallic Au with space group Fm-3m.     

	\emph{Characterization of monolayer MoS$_2$ film.}  CVD-grown MoS\textsubscript{2} monolayer single crystals were purchased from Six Carbon Technology company. The single crystals have triangular shape as observed by optical microscopy (Figure~\ref{fig1} (B)). The SEM (Scanning electron microscope) image of monolayer MoS\textsubscript{2} is shown in the Figure~\ref{fig6} (D). The size of the triangular crystals ranges from 10 $\mu$m to 30 $\mu$m and the average size of crystals is about 20 $\mu$m. The height of MoS$_2$ crystal was estimated using an atomic force microscopy (AFM), Bruker Nano Inc Dimension Icon from Bruker. An AFM image of a flake of MoS$_2$ is shown in Figure~\ref{fig6}. The flake has a height of 8$\pm2$ \AA, indicating that it is a monolayer. 
	
	Raman studies were performed using a custom-made confocal micro-Raman with the 532 nm line of a solid-state laser for excitation in the backscattering geometry. The laser probing spot dimension was 2 µm. Raman spectra were recorded with a spectral resolution of 2 cm$^{-1}$ using a single-stage spectrograph equipped with a CCD array detector. The laser power on specimens was kept below 1 mW, to avoid any laser-induced decomposition. Ultra-low-frequency solid-state notch filters allowed us to measure Raman spectra down to 10 cm$^{-1}$.~\cite{Hinton2019-xf}	Raman spectrum of the sample shown in Figure~\ref{fig1}(E) provide additional evidence of the prevalence of monolayers of MoS$_2$.

\emph{Preparation of hybrid system.} The monolayer MoS\textsubscript{2} sample is first quenched in a vacuum oven at 200$^{\circ}$ C for half an hour, and then gold electrodes are plated on the sample. After the measurements of PC and PL from the pristine sample, Au nanoparticles are dispersed with acetone. , and the gold nanoparticle solution is dripped onto the sample. 

\subsection{Measurement of PC and PL}

\begin{figure}[htb]
	\centering
	\includegraphics[width=0.9\linewidth]{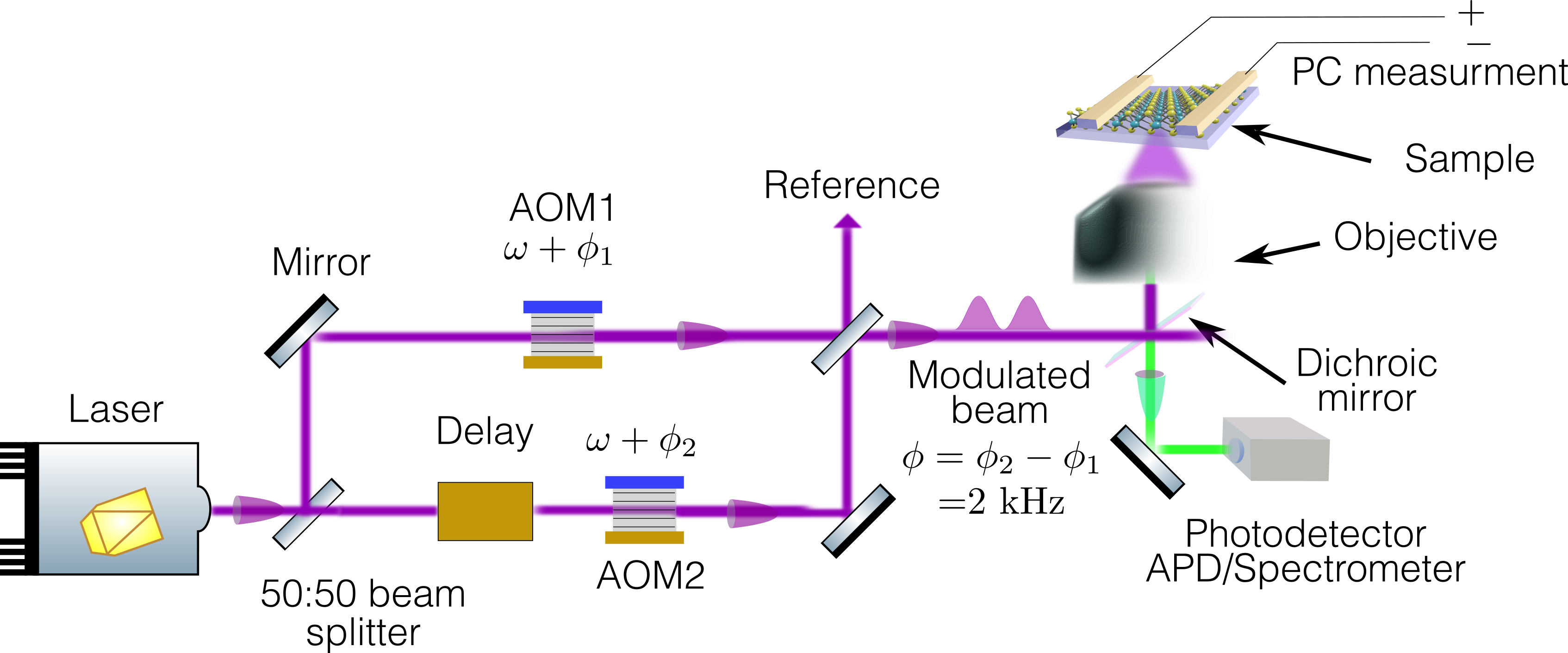}
	\caption{ Experimental setup to measure the photocurrent (PC) and emission from MoS$_2$ using an intensity modulated laser beam. }
	\label{fig_setup}
\end{figure}

 Figure~\ref{fig_setup} shows the schematic of the optical setup used to generate intensity modulated light field. It is a simplification of the four-beam setup that is used in the PC and PL detected multidimensional spectroscopy.\cite{karki2017increasing,karki2014coherent,karki2019before,bian2020vibronic} It uses the interference of two beams in a Mach-Zehnder interferometer. The two beams are obtained from a laser using a 50/50\% beam splitter. Each of the outputs of the splitter pass through an acousto-optic frequency shifter (AOFS). The AOFSs are driven by Radio Frequency (RF) generators. The 1st-order diffractions from the AOFS are selected and combined collinearly by another beam splitter. The temporal overlap of the two beams is adjusted by a piezo-driven delay stage in one of the arms of the interferometer. The output of the interferometer is modulated at the difference of the driving frequencies of the AOFSs, which is set to 2 kHz in the measurements.
 
 The intensity modulated laser beam is directed to an inverted fluorescence microscope (NIKON Eclipse Ti2-U) by a dichroic mirror and focused onto the sample by an objective. The PL from the sample is collected in the epi-direction by the sample objective and separated from the laser beam by the dichroic mirror. Residual scattering of the laser is filtered out by additional  optical filters. The PL is detected by an avalanche photodiode (APD). The signal from the APD is digitized by a data acquisition card.
 
 For the measurement of PC, a bias of 1-3 V is applied to the Au electrodes. The generated PC is preamplified by a current-to-voltage preamplifier. The output of the preamplifier is digitized by the data acquisition card.

\medskip
\textbf{Supporting Information} \par 
Supporting Information is available from the Wiley Online Library or from the author.

\medskip
\textbf{Acknowledgements} \par 
This work was generously supported by National Key R\&D Program of China (2023YFA1407100), Guangdong Science and Technology Department (2021B0301030005, GDZX2304005), and Li Ka Shing Foundation STU-GTIIT
Joint Research Grants (2024LKSFG02).

\medskip

%
\bibliography{MSP-template}

\begin{thebibliography}{10}
\providecommand{\url}[1]{\texttt{#1}}
\expandafter\ifx\csname urlstyle\endcsname\relax
  \providecommand{\doi}[1]{doi:\discretionary{}{}{}#1}\else
  \providecommand{\doi}{doi:\discretionary{}{}{}\begingroup
  \urlstyle{rm}\Url}\fi

\bibitem{Novoselov2004}
K.~S. Novoselov, \emph{et~al.}, Electric Field Effect in Atomically Thin Carbon
  Films. \emph{Science} \textbf{306}, 666--669 (2004),
  \doi{10.1126/science.1102896}.

\bibitem{Xia2014}
F.~Xia, H.~Wang, D.~Xiao, M.~Dubey, A.~Ramasubramaniam, Two-dimensional
  material nanophotonics. \emph{Nature Photonics} \textbf{8}, 899--907 (2014),
  \doi{10.1038/nphoton.2014.271}.

\bibitem{Wang2012}
Q.~H. Wang, K.~Kalantar-Zadeh, A.~Kis, J.~N. Coleman, M.~S. Strano, Electronics
  and optoelectronics of two-dimensional transition metal dichalcogenides.
  \emph{Nature Nanotechnology} \textbf{7}, 699--712 (2012),
  \doi{10.1038/nnano.2012.193}.

\bibitem{Radisavljevic2011}
B.~Radisavljevic, A.~Radenovic, J.~Brivio, V.~Giacometti, A.~Kis, Single-layer
  MoS$_2$ transistors. \emph{Nature Nanotechnology} \textbf{6}, 147--150
  (2011), \doi{10.1038/nnano.2010.279}.

\bibitem{Yu2016}
Y.~Yu, \emph{et~al.}, Engineering Substrate Interactions for High Luminescence
  Efficiency of Transition-Metal Dichalcogenide Monolayers. \emph{Advanced
  Materials} \textbf{28}, 792--797 (2016), \doi{10.1002/adma.201504031}.

\bibitem{Choi2012}
W.~Choi, \emph{et~al.}, High-Detectivity Multilayer MoS$_2$ Phototransistors
  with Spectral Response from Ultraviolet to Infrared. \emph{Advanced
  Materials} \textbf{24}, 5832--5836 (2012), \doi{10.1002/adma.201201781}.

\bibitem{Heinz2017}
T.~F. Heinz, H.~A. Atwater, Nonlinear Light-Matter Interaction at Optical
  Interfaces. \emph{The Journal of Physical Chemistry C} \textbf{121},
  10585--10592 (2017), \doi{10.1021/acs.jpcc.7b02039}.

\bibitem{Wu2018}
R.~Wu, \emph{et~al.}, Plasmon-Enhanced Second-Harmonic Generation in MoS$_2$
  with Plasmonic Nanostructures. \emph{Advanced Materials} \textbf{30}, 1704477
  (2018), \doi{10.1002/adma.201704477}.

\bibitem{Jadidi2016}
M.~M. Jadidi, \emph{et~al.}, Nonlinear Terahertz Absorption of Graphene
  Plasmons. \emph{Nano Letters} \textbf{16}, 2734--2738 (2016),
  \doi{10.1021/acs.nanolett.6b00401}.

\bibitem{Cheng2011}
Y.~Cheng, Z.~Zhu, G.~Huang, U.~Schwingenschlögl, Electronic structure of
  monolayer and bilayer transition metal dichalcogenides MX$_2$ (M = Mo, W; X =
  S, Se) from ab initio theory. \emph{Physical Review B} \textbf{83}, 115449
  (2011), \doi{10.1103/PhysRevB.83.115449}.

\bibitem{Butun2015}
S.~Butun, S.~Tongay, K.~Aydin, Enhanced Light Emission from Large-Area
  Monolayer MoS$_2$ Using Plasmonic Nanodisc Arrays. \emph{Nano Letters}
  \textbf{15}, 2700--2704 (2015), \doi{10.1021/acs.nanolett.5b00407}.

\bibitem{Guo2015}
Z.~Guo, \emph{et~al.}, From Black Phosphorus to Phosphorene: Basic Solvent
  Exfoliation, Evolution of Raman Scattering, and Applications to Ultrafast
  Photonics. \emph{Advanced Functional Materials} \textbf{25}, 6996--7002
  (2015), \doi{10.1002/adfm.201502902}.

\bibitem{Oulton2009}
R.~F. Oulton, \emph{et~al.}, Plasmon lasers at deep subwavelength scale.
  \emph{Nature} \textbf{461}, 629--632 (2009), \doi{10.1038/nature08364}.

\bibitem{Kauranen2012}
M.~Kauranen, A.~V. Zayats, Nonlinear plasmonics. \emph{Nature Photonics}
  \textbf{6}, 737--748 (2012), \doi{10.1038/nphoton.2012.244}.

\bibitem{Lin2023}
S.~Lin, J.~Li, J.~Xu, Plasmon-Enhanced Nonlinear Optical Effects in 2D
  Materials. \emph{Advanced Materials} \textbf{35}, 2206397 (2023),
  \doi{10.1002/adma.202206397}.

\bibitem{GERBI2025}
Z.~D. Gerbi, Electronic and optical properties of molybdenum disulfide ({MoS2})
  mono layer using density functional theory ({DFT}) calculations. \emph{AIP
  Adv.} \textbf{15}~(2) (2025).

\bibitem{LAMBRECHT2012}
T.~Cheiwchanchamnangij, W.~R. Lambrecht, Quasiparticle band structure
  calculation of monolayer, bilayer, and bulk MoS 2. \emph{Physical Review
  B—Condensed Matter and Materials Physics} \textbf{85}~(20), 205302 (2012).

\bibitem{HEINZ2010}
K.~F. Mak, C.~Lee, J.~Hone, J.~Shan, T.~F. Heinz, Atomically thin {MoS$_2$}: a
  new direct-gap semiconductor. \emph{Phys. Rev. Lett.} \textbf{105}~(13),
  136805 (2010).

\bibitem{GIES2015}
A.~Steinhoff, \emph{et~al.}, Efficient excitonic photoluminescence in direct
  and indirect band gap monolayer {MoS2}. \emph{Nano Lett.} \textbf{15}~(10),
  6841--6847 (2015).

\bibitem{KAXIRAS2016}
D.~Kaplan, \emph{et~al.}, Excitation intensity dependence of photoluminescence
  from monolayers of {MoS} 2 and {WS} 2 {/MoS} 2 heterostructures. \emph{2d
  Mater.} \textbf{3}~(1), 015005 (2016).

\bibitem{Li2012}
H.~Li, \emph{et~al.}, From Bulk to Monolayer MoS$_2$: Evolution of Raman
  Scattering. \emph{Adv. Funct. Mater.} \textbf{22}, 1385--1390 (2012),
  \doi{10.1002/adfm.201102111}.

\bibitem{yu2013controlled}
Y.~Yu, \emph{et~al.}, Controlled scalable synthesis of uniform, high-quality
  monolayer and few-layer MoS2 films. \emph{Sci. Rep.} \textbf{3}~(1), 1866
  (2013).

\bibitem{SWAN2017}
J.~W. Christopher, B.~B. Goldberg, A.~K. Swan, Long tailed trions in monolayer
  {MoS$_2$}: Temperature dependent asymmetry and resulting red-shift of trion
  photoluminescence spectra. \emph{Sci. Rep.} \textbf{7}~(1) (2017).

\bibitem{QU2021}
S.~Golovynskyi, \emph{et~al.}, Trion binding energy variation on
  photoluminescence excitation energy and power during direct to indirect
  bandgap crossover in monolayer and few-layer {MoS$_{2}$}. \emph{J. Phys.
  Chem. C Nanomater. Interfaces} \textbf{125}~(32), 17806--17819 (2021).

\bibitem{DHARA2022}
C.~A. Bhuyan, K.~K. Madapu, S.~Dhara, Excitation-dependent photoluminescence
  intensity of monolayer {MoS$_2$}: Role of heat-dissipating area and
  phonon-assisted exciton scattering. \emph{J. Appl. Phys.} \textbf{132}~(20),
  204303 (2022).

\bibitem{karki2015phase}
K.~J. Karki, L.~Kringle, A.~H. Marcus, T.~Pullerits, Phase-synchronous
  detection of coherent and incoherent nonlinear signals. \emph{Journal of
  Optics} \textbf{18}~(1), 015504 (2015).

\bibitem{KARKI2019}
P.~Kumar, Q.~Shi, K.~J. Karki, Enhanced radiative recombination of excitons and
  free charges due to local deformations in the band structure of
  {MAPbBr$_{3}$} perovskite crystals. \emph{J. Phys. Chem. C}
  \textbf{123}~(22), 13444--13450 (2019).

\bibitem{MATHUR2013}
P.~Vasa, \emph{et~al.}, Supercontinuum generation in water doped with gold
  nanoparticles. \emph{Appl. Phys. Lett.} \textbf{103}~(11), 111109 (2013).

\bibitem{KARKI2018}
J.~Chen, \emph{et~al.}, Evidence of high-order nonlinearities in supercontinuum
  white-light generation from a gold nanofilm. \emph{ACS Photonics}
  \textbf{5}~(5), 1927--1932 (2018).

\bibitem{ZAYATS2016}
A.~V. Krasavin, P.~Ginzburg, G.~A. Wurtz, A.~V. Zayats, Nonlocality-driven
  supercontinuum white light generation in plasmonic nanostructures. \emph{Nat.
  Commun.} \textbf{7}~(1), 11497 (2016).

\bibitem{ALBELLA2019}
A.~I. Barreda, J.~M. Saiz, F.~Gonz{\'a}lez, F.~Moreno, P.~Albella, Recent
  advances in high refractive index dielectric nanoantennas: Basics and
  applications. \emph{AIP Adv.} \textbf{9}~(4), 040701 (2019).

\bibitem{MIYAZAKI2016}
N.~Fujimura, A.~Ohta, K.~Makihara, S.~Miyazaki, Evaluation of valence band top
  and electron affinity of {SiO$_2$} and Si-based semiconductors using X-ray
  photoelectron spectroscopy. \emph{Jpn. J. Appl. Phys. (2008)}
  \textbf{55}~(8S2), 08PC06 (2016).

\bibitem{STEMANS2015}
V.~V. Afanas'ev, \emph{et~al.}, Band alignment at interfaces of few-monolayer
  {MoS$_2$} with {SiO$_2$} and {HfO$_2$}. \emph{Microelectron. Eng.}
  \textbf{147}, 294--297 (2015).

\bibitem{HENKER2007}
D.~K{\"o}nig, M.~Rennau, M.~Henker, Direct tunneling effective mass of
  electrons determined by intrinsic charge-up process. \emph{Solid State
  Electron.} \textbf{51}~(5), 650--654 (2007).

\bibitem{SEBASTIAN2009}
R.~S. Swathi, K.~L. Sebastian, Long range resonance energy transfer from a dye
  molecule to graphene has (distance)(-4) dependence. \emph{J. Chem. Phys.}
  \textbf{130}~(8), 086101 (2009).

\bibitem{WRACHTRUP2013}
J.~Tisler, \emph{et~al.}, Single defect center scanning near-field optical
  microscopy on graphene. \emph{Nano Lett.} \textbf{13}~(7), 3152--3156 (2013).

\bibitem{CARUSO2004}
E.~Dulkeith, \emph{et~al.}, Plasmon emission in photoexcited gold
  nanoparticles. \emph{Phys. Rev. B Condens. Matter Mater. Phys.}
  \textbf{70}~(20) (2004).

\bibitem{KIM2024}
H.~T. Nguyen, X.~A. Nguyen, A.~T. Hoang, T.~J. Kim, Spectroscopic ellipsometry
  study of the temperature dependences of the optical and exciton properties of
  MoS$_2$ and WS$_2$ monolayers. \emph{Materials (Basel)} \textbf{17}~(22),
  5455 (2024).

\bibitem{Mathur2021-ml}
D.~Mathur, \emph{et~al.}, Understanding F{\"o}rster resonance energy transfer
  in the sheet regime with {DNA} brick-based dye networks. \emph{ACS Nano}
  \textbf{15}~(10), 16452--16468 (2021).

\bibitem{MULVANEY1996}
L.~M. Liz-Marz{\'a}n, M.~Giersig, P.~Mulvaney, Synthesis of nanosized
  {Gold--Silica} {Core--Shell} particles. \emph{Langmuir} \textbf{12}~(18),
  4329--4335 (1996).

\bibitem{NG2020}
Y.-H. Cheng, K.-M. Ng, Sensitive detection of separated charges in nanohybrids
  by laser excitation mass spectrometry with tetrabutylammonium cationic probe.
  \emph{Anal. Chem.} \textbf{92}~(15), 10262--10267 (2020).

\bibitem{Hinton2019-xf}
J.~K. Hinton, \emph{et~al.}, Effects of pressure on the structure and lattice
  dynamics of $\alpha$-glycine: a combined experimental and theoretical study.
  \emph{CrystEngComm} \textbf{21}~(30), 4457--4464 (2019).

\bibitem{karki2017increasing}
K.~J. Karki, Increasing the density of modes in an optical frequency comb by
  cascaded four-wave mixing in a nonlinear optical fiber. \emph{Physical Review
  A} \textbf{96}~(4), 043802 (2017).

\bibitem{karki2014coherent}
K.~J. Karki, \emph{et~al.}, Coherent two-dimensional photocurrent spectroscopy
  in a PbS quantum dot photocell. \emph{Nature Communications} \textbf{5}~(1),
  5869 (2014).

\bibitem{karki2019before}
K.~J. Karki, \emph{et~al.}, Before F{\"o}rster. Initial excitation in
  photosynthetic light harvesting. \emph{Chemical science} \textbf{10}~(34),
  7923--7928 (2019).

\bibitem{bian2020vibronic}
Q.~Bian, \emph{et~al.}, Vibronic coherence contributes to photocurrent
  generation in organic semiconductor heterojunction diodes. \emph{Nature
  communications} \textbf{11}~(1), 617 (2020).

\end{thebibliography}
\bibliographystyle{sciencemag}

\section*{Supplementary Materials for\\ \scititle}

\subsubsection*{Comparison of the yields of TPL and SHG.}
As shown in the emission spectra in Figure (2) of the main text, TPL is substantially less than SHG.  
The comparatively low detection of PL has two reasons. First, as it is emitted isotropically in all directions, only a fraction of it is collected by the microscope objective. With a 10X objective having a working distance of 16 mm and the pupil diameter of 14 mm, only the fraction emitted over $\sim$ 0.5 sr is collected, which is about 4\% of the total emission. On the other hand, as the SHG has a beam divergence comparable to the laser, and is emitted either in the forward or the backward direction, its collection efficiency is 50\%. Now, the measured yields of SHG and TPL obtained from the area under the corresponding peaks in the spectrum are $I_{SHG}= 5290$ and $I_{PL}=40$ (units proportional to the number of photons), with $I_{PL}/I_{SHG}  \approx 0.008.$ Hence, the ratio of their quantum yields, $K$, is given by:
$K= (I_{SHG}/0.5)/(I_{PL}/0.04)  \approx 11$.
It is important to note that SHG depends on the orientation of the crystal and the PL depends on the density of defects. The ratio varies significantly in different measurements. Nevertheless, in all the samples, when measurements are done by optimizing the SHG output, its yield has been found to be significantly larger than TPL.

\subsubsection*{Simulation of the electric field around the Au nanoparticles}

The electric field distribution around 30 nm Au nanoparticle with 10 nm silica shell on top of a saphhire plate was calculated using the Finite Difference Time Domain (FDTD) method. Parameters used in the simulation are given in table below:

\begin{table}[h]
	\centering
	\caption{Parameters used in the FDTD simulations}
	\label{tab:FDTD}
	\begin{tabular}{lr}
		\toprule
		\textbf{Center wavelength of incident beam} & 1030 nm \\
		\midrule
		Beam propagation direction & Into the plane \\
		\hline
		Beam polarization & P polarized \\
		\hline
		Simulation area & 100$\times$100 nm$^2$ \\
		Mesh distance  & 0.5 nm  \\
		\hline
		Diameter of Au nanoparticle & 30 nm \\
		\hline
		Thickness of silica shell & 10 nm \\
		\hline
		Size of sapphire plate & $80\times80\times80$ nm$^3$\\
		\bottomrule
	\end{tabular}
\end{table}

\begin{figure}[htbp] 
		\centering
		\includegraphics[width=0.9\textwidth]{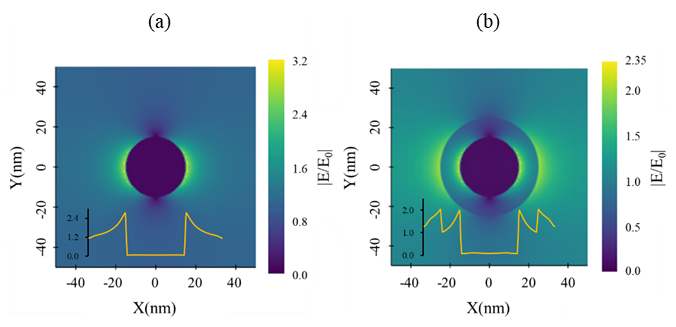} 
		
		\caption{a) The field enhancement around the Au nanoparticle; b) around the Au-core silica shell nanoparticle.}
	\label{S1} 
\end{figure}

	As illustrated in Figure~\ref{S1}(a), the field enhancement around the Au nanoparticle is most pronounced near the particle's surface on the plane through the center of the particle aligning with the direction of polarization. The highest field enhancement observed around the Au nanoparticle is 2.72. Specifically, the line connecting the two points with the strongest enhancement is parallel to the polarization direction of the incident laser, passing through the center of the gold nanoparticle. The field distribution around Au nanoparticles with silica shell is shown in Figure~\ref{S1}(b). The field on the plane parallel to the substrate and passing through the center of the nanoparticle is enhanced at the Au/silica and silica/air interfaces. The enhancement factor at both the interfaces is about 1.9. However, the enhancement of the field is less on the planes closer to the substrate. In fact, the field is suppressed at the bottom of the nanoparticle as discussed in the main text. 

\begin{figure}[htbp] 
		\centering
		\includegraphics[width=0.8\textwidth]{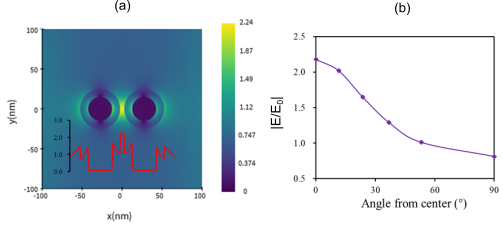} 
		
		\caption{a) The electric field around a pair of Au-core silica shell nanoparticles on the plane through the center of the particles. b) The field on the circumference of one of the particles.  90$^{\circ}$ is at the base of the particle.  }
	\label{S2} 
\end{figure}

As the nanoparticles are dispersed by drop-casting, their aggregation is not controlled. We have also simulated field around a pair of core-shell nanoparticles as an approximation of the enhancement effects by aggregates. The distance between the particles is 5 nm. As shown in Figure~\ref{S2}(a), the field distribution in between the nanoparticles is enhanced by a factor of 2.2, which is larger than in the case of a single particle. However, the enhancement decreases as one moves away from the point of maximum enhancement towards the sapphire substrate. As shown in Figure~\ref{S2}(b), the field at the base of the nanoparticle (at 90$^{\circ}$) is less than the incident field.

\begin{figure}[htbp] 
		\centering
		\includegraphics[width=0.8\textwidth]{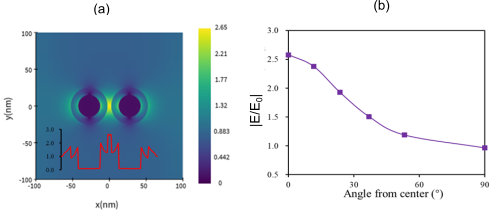} 
		
		\caption{a) The electric field around a pair of Au-core silica shell nanoparticles on the plane through the center of the particles when the incidence angle of the laser beam is at 10$^{\circ}$ to the surface normal. b) The field on the circumference of one of the particles.  90$^{\circ}$ is at the base of the particle.  }
	\label{S3} 
\end{figure}

The simulations show that the maximum enhancement is on the plane perpendicular to the propagation of the incident beam. In our experiments, as the beam is focused on to the sample, it subtends a range of angles determined by the beam diameter and the focal length. The maximum angle, as discussed in the main text, is 4$^{\circ}$. Hence, we have also simulated the enhancement of the field caused by laser beam incident at different angles. Figure~\ref{S3}(a) shows the field distribution around a pair of nanoparticles when the laser incident direction is at 10$^{\circ}$ to the normal of the substrate plane. The maximum field enhancement occurs at the midpoint of the line connecting the centers of the two nanoparticles, with a peak value of 2.58. Figure~\ref{S3}(b) illustrates the field enhancement on the surface of one of the nanoparticles.  The field strength decreases along the nanoparticle surface as one moves toward the silica substrate. At the base of the nanoparticle, the field strength is similar to that of the incident field. Hence, in our experimental condition, we don't expect significant contribution to two-photon absorption by plasmonic enhancement.  

\subsubsection*{Estimation of the energy transfer from Au nanoparticles to MoS$_2$}
The rate of energy transfer ($k_{ET}$) from a point particle to a 2D material depdends on the inverse fourth power of the distance ($d$). This is different from the inverse sixth power ($d^{-6}$) relationship observed for molecule-to-molecule energy transfer. The transfer rate is given by 
\begin{equation}
	k_{ET}(d) = \frac{k_D}{d^4} R_{2D}^4,
\end{equation} 
where $k_D$  is the radiative decay of the donor, and $R_{2D}$ is the characteristic F\"orster distance for the 2D system (see Ref. 35 and 36 in the main text). $R_{2D}$ is a crucial parameter that depends on the optical properties of both the donor and acceptor. It is given by (ref. 36 in the main text)
\begin{equation}
	R_{2D}^4 = \frac{9 e^2\hbar^3 c^3}{4.512 (\hbar \omega)^4 \varepsilon_0 \varepsilon_r},
\end{equation} 
with $\hbar=1.05\times 10^{-34}$ kg/s, $e=1.6\times10^{-9}$ C, $\varepsilon_0=8.854\times10^{-12}$ F/m, $\varepsilon_r$ as the dielectric function of monolayer MoS$_2$. Because the emission is broadband, the characteristic distance varies significantly over the spectrum, ranging from 33 nm to 16 nm. We assume the emitting particles (hot electrons) are distributed evenly on the surface of Au core. Their average distance from the MoS$_2$ flake is about 25 nm. The efficiency of the energy transfer is given by (see Ref. 39) 

\begin{equation}
	\eta = \frac{1}{1+\left(\frac{d}{R_{2D}}\right)^4}.
\end{equation}
The efficiency varies from 70\%  to 20\% for the emissions at 2-3 eV.

\end{document}